\documentclass[12pt,preprint]{aastex}
\shorttitle{X-ray Afterglow Spectra of GRB~020813 and GRB~021004}
\shortauthors{Butler et al.}

\begin{document}

\title{The X-ray Afterglows of GRB~020813 and GRB~021004 with
Chandra HETGS: Possible Evidence for a Supernova Prior to GRB~020813}

\author{Nathaniel R. Butler\altaffilmark{1},
Herman L. Marshall\altaffilmark{1},
George R. Ricker\altaffilmark{1},
Roland K. Vanderspek\altaffilmark{1},
Peter G. Ford\altaffilmark{1},
Geoffrey B. Crew\altaffilmark{1},
Donald Q. Lamb\altaffilmark{2},
J. Garrett Jernigan\altaffilmark{3}}
\altaffiltext{1}{Center for Space Research, Massachusetts Institute of
	Technology, Cambridge, MA 02139}
\altaffiltext{2}{Department of Astronomy \& Astrophysics, University of Chicago,
Chicago, IL 60637}
\altaffiltext{3}{Space Sciences Laboratory, University of California, Berkeley, CA 94720}
\email{nrbutler@space.mit.edu, hermanm@space.mit.edu,
grr@space.mit.edu, roland@space.mit.edu, pgf@space.mit.edu,
gbc@space.mit.edu, lamb@pion.uchicago.edu, jgj@ssl.berkeley.edu}
\slugcomment{Accepted to Ap. J.}

\def\eg{{\it e.g.}}
\newcommand{\ie}{{\it i.e.\,}}
\def\etal{{\it et al.\,}}
\def\et{{\it et al.\,}}
\def\etc{{\it etc.}}
\def\gtrsim{\mathrel{\hbox{\rlap{\hbox{\lower4pt\hbox{$\sim$}}}\hbox{$>$}}}}
\def\lessim{\mathrel{\hbox{\rlap{\hbox{\lower4pt\hbox{$\sim$}}}\hbox{$<$}}}}
\def\arcmin{$^{\prime}$}
\def\arcsec{$^{\prime\prime}$}
\def\degree{$^{\circ}$}
\def\Vsection{\vspace{-0.11in}\section}
\def\Vsubsection{\vspace{-0.11in}\subsection}
\def\Vsubsubsection{\vspace{-0.11in}\subsubsection}
\newcommand\hete{{\it HETE}}
\newcommand\xmm{{\it XMM-Newton}}
\newcommand\chandra{{\it Chandra}}

\begin{abstract}
We report on the detection of an emission line near 1.3 keV,
which we associate with blue-shifted hydrogen-like sulfur (S~XVI),
in a 76.8 ksec Chandra HETGS spectrum of the afterglow of
GRB~020813.  The line is detected at $3.3\sigma$ significance.
We also find marginal evidence for a line possibly due to
hydrogen-like silicon (Si~XIV) with the same blue-shift. 
A line from Fe is not detected, though a very low significance Ni
feature may be present.  A thermal model fits the data adequately, but 
a reflection model may provide
a better fit.   There is marginal evidence that the equivalent width of
the S~XVI line decrease as the burst fades. 
We infer from these results that a supernova likely
occurred $\gtrsim$ 2 months prior to the GRB.  We find no discrete or 
variable spectral features in the Chandra HETGS spectrum of the 
GRB~021004 afterglow.
\end{abstract}


\section{Introduction}
\label{sec:intro}

Much evidence has accumulated in recent years connecting
long-duration $\gamma$-ray bursts (GRBs) \citep{kouv93} to 
supernovae (SNe).  The detection of
SN 1998bw in association with GRB~980425 \citep{galama98}
and the detection of 
late-time SN ``bumps'' in several afterglow light curves (\citet{bloom99}, 
\citet{reichart01} and references therein), the observed
positional coincidences 
between star-forming regions and long-duration GRB afterglows (e.g.
\citet{sahu97}, \citet{kulkarni98}, \citet{kulkarni99}), and the
evidence for dust extinction in some GRB afterglows all point toward
this association.  In this context, X-ray 
spectroscopic observations of early GRB afterglows
are tremendously interesting, because they potentially allow us to view
directly the effects of the GRB on the progenitor stellar material and
possibly on metals freshly synthesized in the SN.

Several authors have claimed detection of Fe lines in GRB X-ray 
afterglows (\citet{piro99}, \citet{piro00}, \citet{antonelli00}).
Beyond very weak evidence in \citet{piro00} for an emission line from
H-like S in GRB~991216, \citet{reeves02} (hereafter R02) are the first 
to strongly assert evidence for emission from low-Z, multiple-$\alpha$
elements in a GRB 
X-ray afterglow.  If valid, this claim has broad implications for GRB
emission models and it would strongly link GRBs to SNe.
The reduction of these data and the statistical significance of the
results has been called into question by \citet{boroz02} and
\citet{rutledge02}, respectively.  In a follow-up paper \citep{reeves02b}, 
the \xmm~(\citep{jansen01}) observers address those concerns.  Physical 
interpretations aside,
the R02 results are controversial largely due to the small
number of observed counts and the low spectral resolution of the EPIC-pn
instrument, which would blur some fraction of the line emission into
the continuum.  These hinder an accurate measurement of the continuum 
emission and make it difficult to gauge the significance of discrete
spectral features.

We find possible evidence for Si and S 
emission in the {\it Chandra X-ray Observatory} \citep{weisskopf02} 
High Energy Transmission Grating Spectrometer (HETGS) spectrum of 
the GRB~020813 afterglow.  We detect a Ni-K line at very low 
significance, and we find a strikingly low upper limit to the Fe-K line 
equivalent width.  The emitting material is blue-shifted with velocity 
$\sim 0.1c$ relative to the GRB host galaxy. 
These results add weight to the detections claimed by R02 and \citet{watson02}.
With the HETGS's high spectral resolution, we are able to resolve the line
widths, and we are able to accurately separate the line and continuum emission.
Because the lines we associate with S and Si are detected in the full data set, 
we have $\sim 5$ times more counts to work with than R02, where the line
emission is observed during only a short period of time.  Moreover, the 
continuum is an order of magnitude stronger than that reported by R02, 
possibly indicating, in the case of thermal emission, that a power-law 
component is also present.  This may alternatively be a sign that the lines 
are produced under reflection.

\section{Observations}
\label{sec:observe}

Using Directors Discretionary Time, two afterglows
of GRBs detected and localized by the {\it High-Energy 
Transient Explorer} (\hete) satellite
\citep{ricker03} were observed with the \chandra~HETGS, approximately
1 day after each burst, and lasting for approximately 1 day.  
Both bursts were long-duration GRBs, with $\gamma-$ray
durations $\gtrsim$100s.  The $\gamma-$ray peak flux from GRB~020813 
\citep{villasenor02} was
quite high ($\sim 10^{-5}$ erg cm$^{-2}$ s$^{-1}$), 
while the peak flux from GRB~021004 \citep{shirasaki02} was approximately two
orders of magnitude lower. 

\chandra~began observing the GRB~020813 afterglow with the HETGS 
on August 13.990 ($t_{burst}$ + 21.02 hrs)
and continued until August 14.892 ($t_{burst}$ + 42.67 hrs), with a total
livetime of 76.8 ksec.  The mean count rate from the HEG 
and MEG gratings and the 0th-order image was 0.12 counts/s (0.06 counts/s
for the gratings alone).  
The afterglow candidate reported by \citet{fox0813}
was detected and was observed to fade in brightness with time according
to a power-law of index $\alpha = -1.38 \pm 0.06$ ($\chi^2_{\nu} = 36.4/24$), 
consistent with the value reported by \citet{vanderspek02}
and consistent with measured values in the optical \citep{malesani02}.  Between
0.6 and 6 keV, we measure a time-averaged flux of $2.2 \times 10^{-12}$ erg 
cm$^{-2}$ s$^{-1}$.

As first reported by \citet{sako02}, \chandra~began observing 
the afterglow of GRB~021004 on October 5.358 ($t_{burst}$ + 20.49 hrs) 
and continued until October 6.400 ($t_{burst}$ + 45.50 hrs), with a total 
livetime of 86.7 ksec.  The mean count rate from the HEG and MEG gratings 
and the 0th-order image was 0.04 counts/s (0.02 counts/s in the gratings
alone).  The afterglow candidate reported 
by \citet{fox1004} was detected and was observed to fade in brightness
with time.  The X-ray light curve appears variable.  (Similar variability
has been reported in the optical for this afterglow; see e.g. 
\citet{halpern02}).  The temporal fade in flux can be described by a 
power-law with index $\alpha = -0.9 \pm 0.1$ ($\chi^2_{\nu} = 83.8/27$), 
consistent with the values reported by \citet{sako02} and \citet{holland02}.  
The time-averaged flux between 0.6 and 6 keV is measured to be 
$6.3 \times 10^{-13}$ erg cm$^{-2}$ s$^{-1}$.

The total number of counts detected by
\chandra~in the case of GRB~020813 (GRB~021004) is $6.5$
times (2 times) the number of counts detected for GRB~991216
\citep{piro00}.  We perform a parallel spectral/temporal analysis below 
for the X-ray afterglow spectra of GRB~020813 and GRB~021004.

\section{Spectral Fitting}
\label{sec:spectral_fitting}

We reduce the HETGS spectral data from the standard L2 event lists using 
IDL and custom scripts, described in \citet{marshall02}.  For the 
0th-order spectral data, we use the 
CIAO\footnote{http://cxc.harvard.edu/ciao/}
processing exclusively.  Each data set is corrected for 
QE degradation\footnote{http://asc.harvard.edu/cal/Acis/Cal\_prods/qeDeg/}
due to contamination in the ACIS chips,
prior to spectral fitting.  We also destreak the S4 
chip\footnote{http://asc.harvard.edu/ciao/threads/spectra\_hetgacis/}.
Spectral fitting and analysis is performed with 
ISIS\footnote{http://space.mit.edu/CXC/ISIS/}.  For each 
observation, 
we fit the HETGS HEG and HETGS MEG 1st-order data for the entire observation 
jointly.  The +1 and -1 orders are combined for each grating.  The data are 
then binned to a S/N $\ge 5$ per bin, with the bin width restricted to $\delta 
E / E \le 0.1$.  We define S/N as the background-subtracted number of
counts divided by the square root of the sum of the signal counts and the
variance in the background.   
Bins of the maximal width with S/N $<3$ are rejected, due to lack of
signal.  For GRB~020813, this selects 
an HEG energy range of $0.8-7.6$ keV and and MEG energy 
range of $0.6-5.2$ keV.  For GRB~021004, this selects an HEG energy range of 
$1.0-4.6$ keV and and MEG energy range of $0.6-4.7$ keV.  
We fit each
model by minimizing $\chi^2$.  Unless otherwise noted, all quoted errors are 
90\% confidence.  

\subsection{GRB~020813}

As reported by \citet{vanderspek02}, we find that an absorbed power-law 
fit to the 1st-order 
spectra requires no absorption column in excess of Galactic value.
Therefore, in each model that we consider, we include absorption with 
$N_{\rm H}$ frozen to $7.5 \times 10^{20}$ cm$^{-2}$.

\begin{figure*}
\begin{center}
\plotone{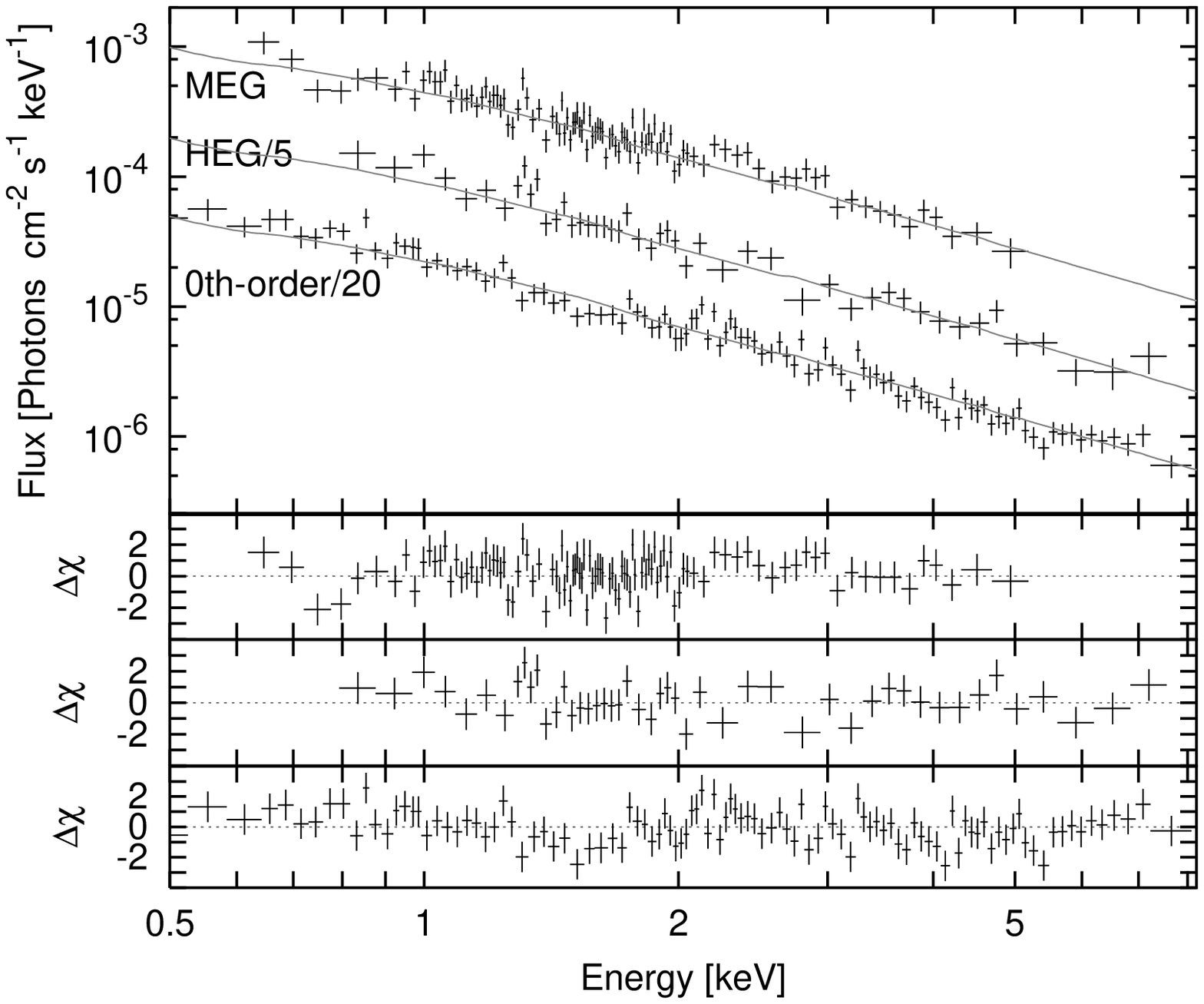}
\caption{
The $\pm1$ order HETGS (HEG/MEG) data for GRB~020813
are moderately well fit by an absorbed 
power-law (Section \ref{sec:spectral_fitting}).  The 0th-order data 
(S/N $\ge 5$ per bin) are also fit well by this model.
Here we divide the HEG data by a factor of 5 and the 0th-order data
by a factor of 20 for ease of viewing.
}
\label{fig:813spec}
\end{center}
\end{figure*}

The data are moderately well fit by a power-law (Figure \ref{fig:813spec}, 
$\chi^2_{\nu}=176.0/154$), with photon index $\Gamma = 1.85 \pm 0.04$ and 
normalization $(5.5\pm 0.2) \times 10^{-4}$
photons/keV/cm$^2$/s at 1 keV. 
A thermal bremsstrahlung fit is a much poorer fit to 
the data ($\chi^2_{\nu}=196.7/154$, rejectable at 99\% confidence), with 
$kT = 4.6 \pm 0.5$ keV.  We find consistent fits using the 0th-order data.  
Of note, a large fraction of the contribution to $\chi^2$ in the case of 
the HETGS power-law fit arises between $1.2$ and $1.5$ keV.  The addition of a 
Gaussian 
component in this energy range improves the $\chi^2$ ($\chi^2_{\nu}=
160.5/151$).  This improvement is at the 99.9\% confidence level
according the the likelihood ratio test ($\Delta\chi^2=15.5$ for 3 additional
degrees of freedom).  However, this method for determining significance is
deprecated (see \citet{protassov02}), and we present Monte Carlo calculations
for the significance below.  The line energy is $1.31 \pm 0.01$ keV, with a 
width of $8\pm 4$ eV (1$\sigma$). The rest-frame equivalent width is
$67\pm 32$ eV, and the flux is $(1.6 \pm 0.8) \times 10^{-14}$ erg 
cm$^{-2}$ s$^{-1}$.  Given the optically measured GRB host galaxy absorption 
redshift $z\geq 1.254$ \citep{price02}, a 
likely association for this line is S XVI K$\alpha$, blue-shifted by 
$0.12c$ from $z=1.254$.  The $\chi^2$ is marginally improved 
($\chi^2_{\nu}=155.7/150$) with the inclusion of an additional Gaussian 
component of the same width, constrained to lie at predicted energy of 
Si XIV K$\alpha$ ($1.01$ keV at $z=0.99 \pm 0.01$).  From the likelihood
ratio test, the confidence for the pair can be estimated as 99.96\% 
($\Delta\chi^2=20.3$ for 4 additional degrees of freedom).
Allowing the widths to vary jointly, the best fit width value is $10\pm 9$ eV.  

\begin{figure*}
\begin{center}
\epsscale{0.8}
\plotone{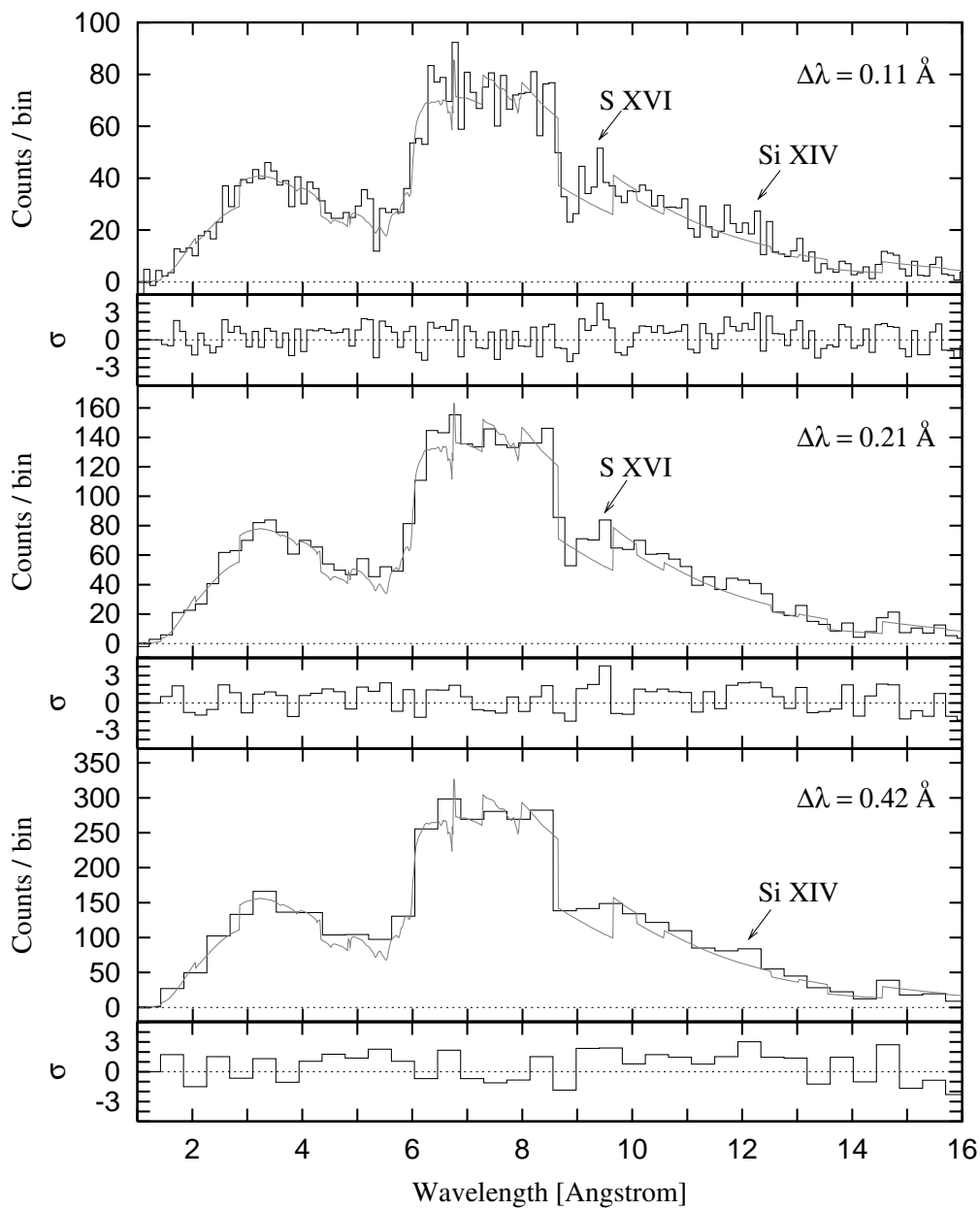}
\caption{
The best fit power-law model on top of the combined HETGS 
HEG+MEG data, at
3 binnings, for GRB~020813.  The significance (reported
in $\sigma$ units) of deviations
from the model are calculated using Poisson statistics.  We associate
the residuals near 9.4 \AA~in the top and middle plots
with H-like S.  We associate the high bin near 12.1 \AA~in the
top and bottom plot with H-like Si.
}
\label{fig:allcts_813}
\end{center}
\end{figure*}

In gauging the significance of these features, we must take into account
the number of trials performed (i.e. line widths and line energies searched).
Figure \ref{fig:allcts_813} displays the 
power-law model found above over-plotted on the summed HEG+MEG counts, at
three binnings.  The finest binning is chosen so that each bin between 1.5 and
16.0 \AA~contains one or more counts.  We rebin twice by a factor of 
approximately two in order to look for broader features.  The $1.31$ keV 
(9.4 \AA) feature is 
seen in both the top and middle plots at $4.0 \sigma$.  This significance 
describes the Poisson
probability that the signal plus background counts could reach a value
greater than or equal to the observed values, with the Poisson mean fixed
to the number of counts predicted by the model plus the background.  The 
probability for a chance fluctuation of 
this or greater significance in one or more of the 235 bins shown in 
Figure \ref{fig:allcts_813} is approximately 1.5\%, 
calculated as $P_N\approx 1-(1-P_1)^N$.  From Monte Carlo simulations of
$10^4$ spectra using the best fit power-law model above, we find
$P_N=1.4\%$.  From the simulations, we find that the probability for
getting two or more 4$\sigma$ fluctuations (as observed) is 0.1\%.
Thus the line is detected at $3.3\sigma$ significance.
A broad feature corresponding to 
the $1.01$ keV line mentioned above is observed in the third panel (and
also in the 1st panel) of Figure \ref{fig:allcts_813} at $\sim$ 12.2 \AA, 
with a significance of $2.9 \sigma$.  We estimate a 58\% probability 
that a $2.9 \sigma$ or greater fluctuation would occur by chance in one or 
more of the trials.  From the Monte Carlo simulations, the probability for
getting two or more $2.9\sigma$ fluctuations is 23\%.  Thus the
line is detected at very low ($1.2\sigma$) significance in a blind
search, and it is potentially meaningful only if the line location
is constrained as in the preceding paragraph.
No additional features at or above $2.9 \sigma$ are 
present in the HETGS spectrum at the binnings searched.
Finally, we derive a $3\sigma$ upper limit of 20 eV for the equivalent 
widths of any
lines unresolved in this search (FWHM $\ll$ 0.11 \AA) between
6 and 16 \AA.  This figure grows $\propto\lambda^{-3}$ shortward of 6 \AA. 

We also attempt to fit these data using a collisional ionization
equilibrium (CIE) plasma model, as has 
been done previously in the case of several recent GRB afterglows 
observed with {\it XMM-Newton} (R02, \citet{watson02}).
We form the model using the Astrophysical Plasma Emission Database (APED) 
accessible through ISIS,  
with the redshift fixed at $z=0.99$.  We utilize a turbulence velocity 
$v_{\rm T}$, fixed at $2000$ km/s, and thermal line profiles to produce the 
best-fit line widths above.   We fix the abundance of metals lighter 
than Mg to solar.  
We tie the Si abundance to the S abundance, and allow Ni to vary freely.
All other metals are frozen to zero abundance.  The CIE plasma fit (Figure 
\ref{fig:813fits}, $\chi^2_{\nu}=177.8/152$) is slightly poorer than the
pure power-law fit.  Allowing the APED metal abundances previously set to
zero to vary 
does not improve the fit.  The best fit plasma temperature is 
$9.2 \pm 1.2$ keV.  The best fit S,Si abundance is 2.9 times solar, and
the linked abundance is greater than solar at 98\% confidence 
($\Delta \chi^2 = 5.5$ for 1 additional parameter).
The best fit Ni abundance is 1.8 times solar.
There is a low ($1.6\sigma$) significance detection of a Ni-K line in the 
MEG spectrum.  This features appears as two bins modestly over the 
power-law fit in Figures \ref{fig:813spec}, \ref{fig:813fits} near 4 keV.  
The APED model fits these bins with a blend of K lines from H and He-like Ni.
The APED model with an abundance of Fe comparable to the Si and S abundances
would produce a very strong and easily detectable Fe-K feature (Figure 
\ref{fig:813fits} Inset).  In Table \ref{table:abund}, we report 1-parameter 
90\% confidence upper limits to the abundances of several elements, with 
upper limits to the line fluxes and rest-frame equivalent widths of H-like 
K lines in each species.  In Figures \ref{fig:813abund}, 
\ref{fig:813fluxes}, and \ref{fig:813eqwidth} we compare the abundances, 
line luminosities, and rest-frame equivalent
widths, respectively, to those reported for other GRB X-ray afterglows.  

Since, the abundances of $\alpha$-particle nuclei produced by
core-collapse SNe are expected to be approximately the same 
(see e.g. \citet{rauscher02}), relative to solar,
the low relative upper limits we find for the abundances of Ar and Ca 
possibly indicate that the Si and S lines are not the result of thermal 
emission from a hot plasma.  After radioactive decay, we
would expect an Fe abundance similar to the low-Z, multiple-$\alpha$ metal 
abundances.  This implies that the line fluxes from one or more of the Fe 
group elements (Fe, Ni, Co) should be higher than we observe.  We discuss 
below the possibility that the lines are due to reflection.  An alternative
possibility, suggested in part by the APED model's apparent tendency
to under-fit the continuum below $\sim$ 1.5 keV (Figure \ref{fig:813spec}), 
is that we must also include a power-law component.  It is then
possible to use the APED model with a lower temperature ($\sim 2$ keV), 
and this weakens the high energy continuum and the lines from Ar, 
Ca, and the Fe group elements relative to the Si and S lines.  We find
models with a single abundance for the $\alpha$-particle nuclei and Fe
which improve upon the power-law fit but do not appear to be as good as 
the power-law plus two Gaussian fit.  Finally, we note that the 
abundances quoted in Table \ref{table:abund} lose meaning if a 
power-law component is also present.

\subsection{GRB~021004}

An absorbed power-law fit to the 1st-order spectra does not require an 
absorption column in excess of the Galactic value, as reported by 
\citet{sako02}.  
Therefore, in each model that we consider, we include absorption with 
$N_{\rm H}$ frozen to $4.24 \times 10^{20}$ cm$^{-2}$.

The data are very well fit by a power-law (Figure \ref{fig:1004spec}, 
$\chi^2_{\nu}=39.5/55$), with $\Gamma = 2.01 \pm 0.08$ and normalization 
$(1.7\pm 0.1) \times 10^{-4}$ photons/keV/cm$^2$/s at 1 keV, consistent with 
the values found by \citet{sako02} and \citet{holland02}.  A 
thermal bremsstrahlung fit is a somewhat poorer, though still very acceptable, 
fit ($\chi^2_{\nu}=47.2/55$), with kT $= 3.2 \pm 0.4$ keV.  We 
find consistent fits using the 0th-order data.  

We plot the combined HEG/MEG data in count space with the power-law 
model over-plotted in Figure \ref{fig:allcts_1004}, for the same binnings 
as in Figure \ref{fig:allcts_813}.  No significant features are 
observed.  The largest fluctuation occurs in the middle panel, where one of 
the $\sim 70$ bins reaches $2.6 \sigma$.  There is approximately an 89\% 
chance that this would happen by chance in one or more of the bins in that 
plot.  At the $3\sigma$ level, this search is sensitive to moderately broad
(FWHM $\sim$ $\delta \lambda$) 
emission and absorption lines between 6 and 16 \AA~with equivalent
widths greater than 50 eV.  Over this same wavelength interval, we 
derive a $3\sigma$ 
upper limit of 35 eV to the equivalent widths of any narrow lines
(FWHM $\ll$ 0.11 \AA).  These figures grow $\propto\lambda^{-3}$
shortward of 6 \AA.  Considering the GRB
redshift of 2.328 determined in the optical by \citet{mirabel02},
several K-lines from S, Ar, Ca, Fe, Co, and Ni would have been detected 
were their equivalent widths larger than these estimated limits.

\section{Temporal Analysis}
\label{sec:temporal_analysis}

As displayed in Figure \ref{fig:indxNnorm}, the afterglow continua, though 
moderately variable in the case of GRB~021004, do not appear to evolve 
spectrally.  To search for time-variable discrete features,
we divide the summed HEG+MEG data sets into 2, 4, or 8
portions of equal numbers of counts.   For each time section,
we examine the counts data at 3 binnings: 0.11, 0.21, and 0.42 \AA~(as in
Figures \ref{fig:allcts_813}, \ref{fig:allcts_1004}). 
For each GRB afterglow, we fit absorbed power-law's to each of these 
42 data sets by minimizing the Cash statistic \citep{cash79}.
We then search the bins between 1 and 16 \AA~for deviations from the best fit
model, interpreting the deviations using Poisson statistics 
as in Section \ref{sec:spectral_fitting}.  We set a limiting detection
threshold of $3.7 \sigma$ so that fewer than one chance detection is
expected in $15 \times 235$ bins searched.  The search yields no
emission or absorption line detections in the case of GRB~021004, while 
for GRB~020813 only
the Si and S lines (Section \ref{sec:spectral_fitting}) are detected.
The Si XIV K$\alpha$ line is detected at $3.7 \sigma$ 
($P_N\approx 53\%$) in
the second quarter (12-28.5 ksec) of data,
while the S XVI K$\alpha$ line (3 detections) peaks at $4.3 \sigma$ 
($P_N\approx 6\%$) in the first quarter (12 ksec) of data.
In Figure \ref{fig:813line}, we plot the S XVI K$\alpha$ line flux 
(measured with Gaussian fits of fixed width and location) and the continuum 
flux versus time.  Using the likelihood ratio test, we find evidence for a 
decrease in equivalent width between the first 28.5 ksec 
of data and the remainder of the observation at 97\% confidence ($\Delta 
\chi^2 = 4.7$ for 1 additional parameter).

\section{Discussion}
\label{sec:discuss}

The most significant discrete feature we observe in afterglow spectrum of 
GRB~020813 is clearly the line we associate with S, which we detect at
$3.3\sigma$ significance.  This is a somewhat marginal detection.
However, the reality of the line is supported by the presence of line
counts in various times slices throughout
the observation and in each of the four independent 1st-order 
spectra (MEG/HEG $\pm1$ orders).  
For the full observation no detector anomalies (e.g. flares, flickering 
pixels) are observed in the ACIS-S CCD regions yielding the HETGS MEG/HEG 
$\pm1$ order counts from this line.  The background is negligible between
0.6 and 6 keV, contributing only 3\% of the total counts.  Pileup in the 
grating spectrum is not a concern for this observation due to the very low 
count rate.  Excess counts corresponding to the line are detected in each 
independent spectrum, and
the numbers are consistent at the $1.5\sigma$ level.  The combined 1st-order
spectra (MEG/HEG) show a consistent number of excess counts at the 
$1\sigma$ level.  The validity of this line detection is possibly strengthened
by an association with S~XVI~K$\alpha$ and by a marginal 
detection of a line at the expected energy for Si~XIV~K$\alpha$.
We have performed a parallel reduction for the GRB~021004 
afterglow data, and this has yielded no convincing evidence for discrete 
spectral features.

Although we find possible evidence for temporal variability in the line
emission for GRB~020813 (Section \ref{sec:temporal_analysis}), the lines
are detected in the full 76.8 ksec observation, and the full data set is
moderately well fit by a plasma in collisional ionization equilibrium
(Section \ref{sec:spectral_fitting}).  Figures 
\ref{fig:813abund}-\ref{fig:813eqwidth} show the line and metallicity 
parameters we derive in relation to those reported for other GRBs in
the literature.  Our required abundances are only modestly super-solar
(Section \ref{sec:spectral_fitting}), in contrast to those reported for several
other GRBs.  The line luminosities are quite low, and the rest-frame equivalent
widths are an order of magnitude smaller than those reported for any
other GRB.  Figure \ref{fig:813line} shows the GRB~020813 continuum
and S~XVI line fluxes versus time alongside the data for GRB~011211 from
R02.  In comparison to GRB~011211, it appears that the longevity of
the line emission (and also the narrowness of the lines) was critical to 
their detection, whereas the 10 times stronger continuum could easily 
have hidden less persistent emission.

We can interpret the longer duration and later onset of the lines
for GRB~020813 relative to GRB~011211 (Figure \ref{fig:813line}) as
a purely geometrical effect due to observing emitting material residing
at a larger radius $R$ from the GRB.
For radially expanding emitting material heated by GRB photons or by
the GRB shock, the narrowness of the lines 
requires that the emission comes from only a small solid angle ($\Omega 
\lessim 0.4$), as viewed from the GRB, consistent with GRB beaming models 
\citep{frail01}.  The line duration
($t_{\rm line} \gtrsim 76.8$ ksec) can be attributed to time delays across
a thin shell, and this leads to the requirement
$R = {ct_{\rm line} \over 1+z}{2\pi \over \Omega} \gtrsim 10^{16}$ cm.
An estimate $R \lessim 10^{18}$ cm for the maximum likely $R$ is determined 
using the smallest solid angle found for a GRB jet in \citet{frail01}.
For the reprocessing material to lie this far out from the GRB, it must
have gotten there prior to the GRB.  The line identifications we have
made (Section \ref{sec:spectral_fitting}) require a blue-shift 
$\sim 0.1c$, and this high outflow velocity implicates expanding
SN ejecta rather than a stellar wind.
The SN would have gone off 0.1 to 10 years
prior to the GRB.  The need for a precursor SN can be avoided if the GRB 
host (absorption) redshift
is taken as a lower limit on the redshift, rather than the actual
redshift.  In that case, an identification of the S~XVI and Si~XIV lines with
Ar~XVIII and S~XVI ($z=1.53$, no blue-shift), respectively, could be made.

In the case of
GRB~011211, R02 assert that a shell of thermal plasma at $R \sim 10^{15}$ cm
can account for the afterglow spectrum and for the duration of the reported 
emission lines.  Taking into account radiation transfer across the shell,
\citet{lazzati02a} argues that a larger radius is required ($R 
\gtrsim 10^{17}$ cm), because the large afterglow luminosity
implies a high density, which implies a cooling time much shorter than
the observed line duration.  Following \citet{lazzati02a}, we find
that a thermal model for GRB~020813 also implies $R \gtrsim 10^{17}$ cm,
except for the case (discussed below) where the luminosity of the thermal 
component is
actually overwhelmed by a ($\sim 10$ times brighter) power-law continuum 
component.  In this case, a solution with $R\sim 10^{16}$ cm is possible. 
The density is low ($n\sim 10^8$ cm$^{-3}$), and this implies
a shell with thickness $dR \sim R/10$ for $\tau_{\rm T}=0.1$.  At larger
radii, the shell must be either very thin or very clumped for the contained 
mass to be similar to that expected from SNe ($M_{\rm ejecta}\lessim 
20 M_{\odot}$, \citet{woosely95}).

\begin{figure*}
\begin{center}
\epsscale{0.9}
\plotone{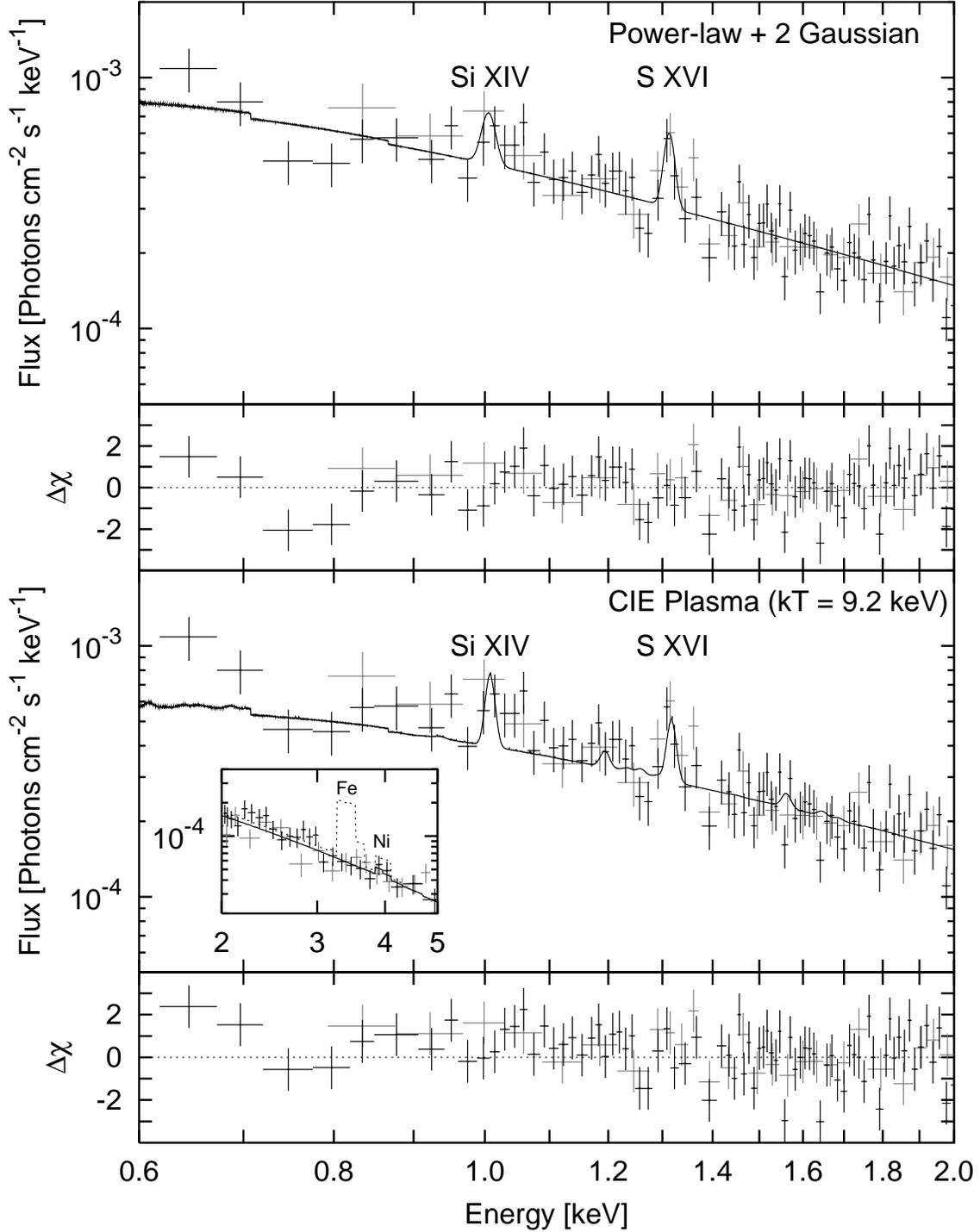}
\caption{
The power-law fit to the GRB~020813 $\pm1$ order HETGS data is improved 
by adding two Gaussian's at the engergies of blue-shifted (by 0.12c)
S~XVI~K$\alpha$ and Si~XIV~K$\alpha$.  The lines and continuum
can be modelled less effectively by a CIE plasma model (bottom plot) containing
Si, S, and Ni.  The inset panel in the bottom plot shows the 2-5 keV portion
of the CIE plasma model fit with (dotted line) and without (1.8 times 
solar abundance in)
Fe; an abundance of Fe similar to the abundances
for Si and S is strongly ruled out.  The HEG data are plotted in grey.
The MEG data are plotted in black.
}
\label{fig:813fits}
\end{center}
\end{figure*}

\begin{figure*}
\begin{center}
\plotone{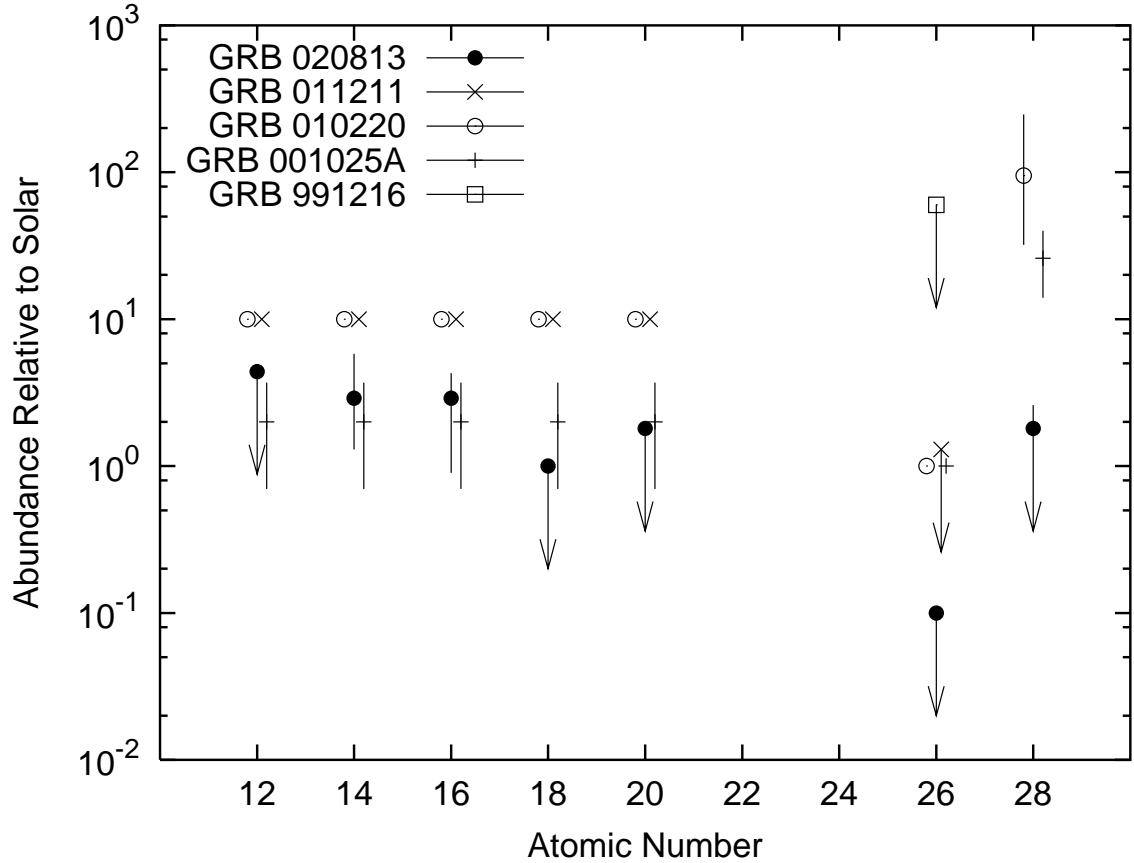}
\caption{
Metal abundances for the APED plasma model for GRB~020813 (solid circles),
as in Table \ref{table:abund}.  Also plotted are the abundances reported
in the literature for several other GRB X-ray afterglow spectra.  
The values for GRB~011211 come from R02. The values for 
GRB~010220 and GRB~001025A come from \citet{watson02}.  The value for
GRB~991216 has been taken from \citet{piro00}.  Error bars have been plotted
where available.
}
\label{fig:813abund}
\end{center}
\end{figure*}

\begin{figure*}
\begin{center}
\plotone{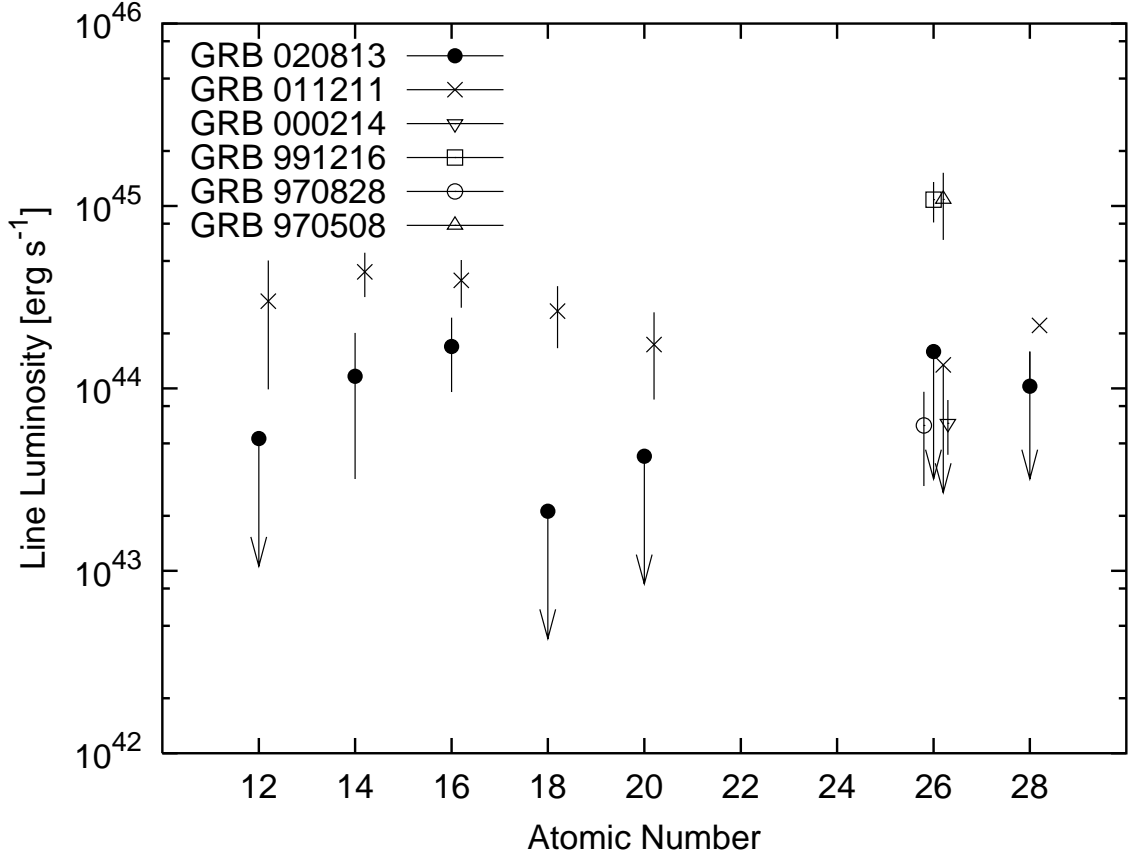}
\caption{
Isotropic equivalent line luminosities from Gaussian fits (Table 
\ref{table:abund}) for GRB~020813 (solid circles, host at $z=1.254$) 
and for several other GRBs in the 
literature.  To determine luminosity distances, we use a cosmology
with $\Omega_m=0.3$, $\Omega_{\Lambda}=0.7$, and $h=0.65$.
The values for GRB~011211 come from R02, with $z=2.14$. 
The value for GRB~000214 comes from \citet{antonelli00}, with $z=0.47$.
The value for GRB~991216 has been taken from \citet{piro00}, using
$z=1$.  The value
for GRB~970828 comes from \citet{yoshida99}, using $z=0.33$.  The value for
GRB~970508 comes from \citet{piro99}, using $z=0.835$.  
Error bars have been plotted where available.
}
\label{fig:813fluxes}
\end{center}
\end{figure*}

\begin{figure*}
\begin{center}
\plotone{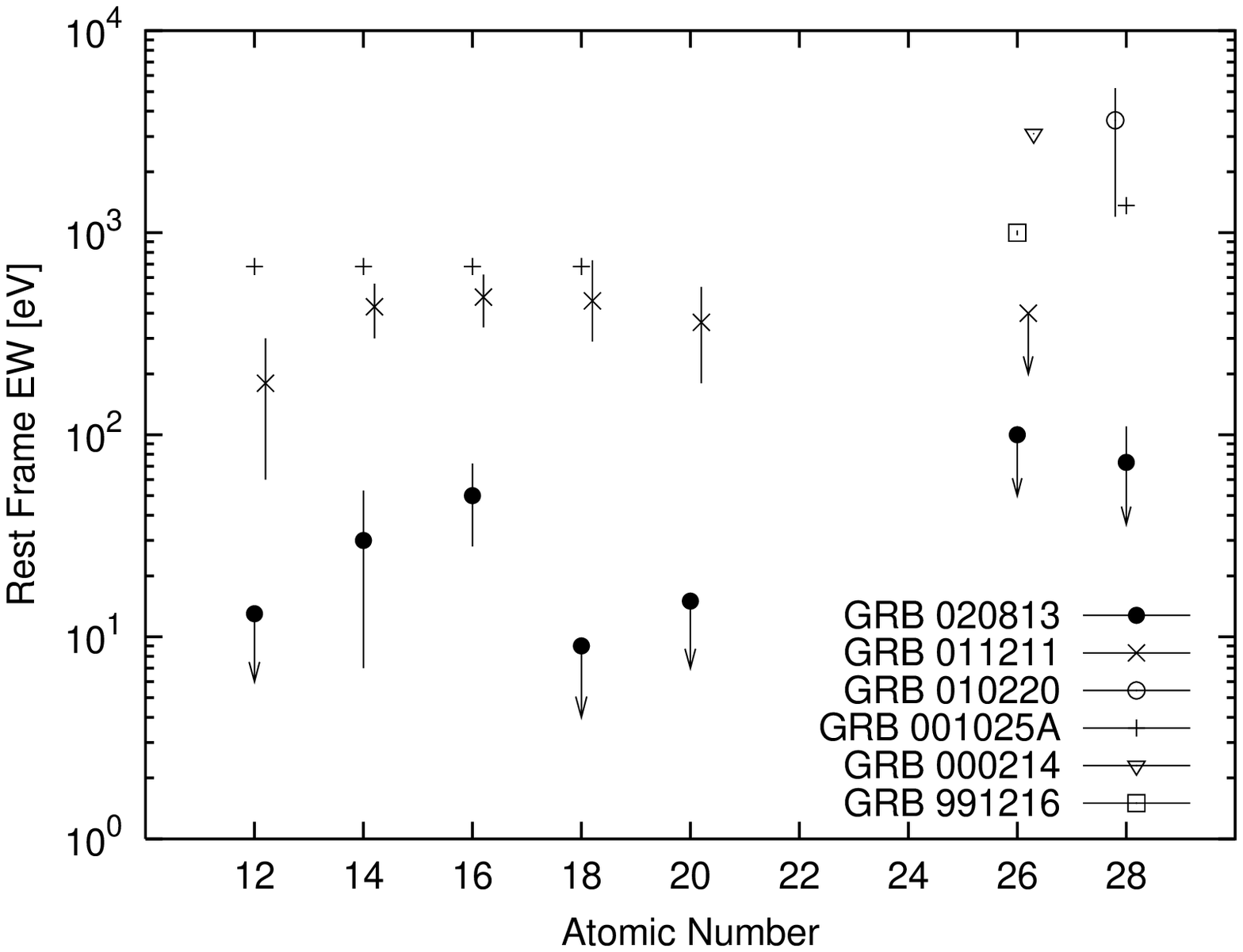}
\caption{
Rest-frame equivalent widths in eV from Gaussian fits (Table \ref{table:abund})
for GRB~020813 (solid circles, $z=0.99$) and for several other
GRBs in the literature.  
The values for GRB~011211 come from R02, with $z=1.88$.
The values for
GRB~010220 ($z=1.0$) and GRB~001025A ($z=0.7$) come from \citet{watson02}.
The value for GRB~000214 comes from \citet{antonelli00}, with $z=0.47$.
The value for GRB~991216 has been taken from \citet{piro00}, using
$z=1$.  Error bars have been plotted where available.
}
\label{fig:813eqwidth}
\end{center}
\end{figure*}

\begin{figure*}
\begin{center}
\plotone{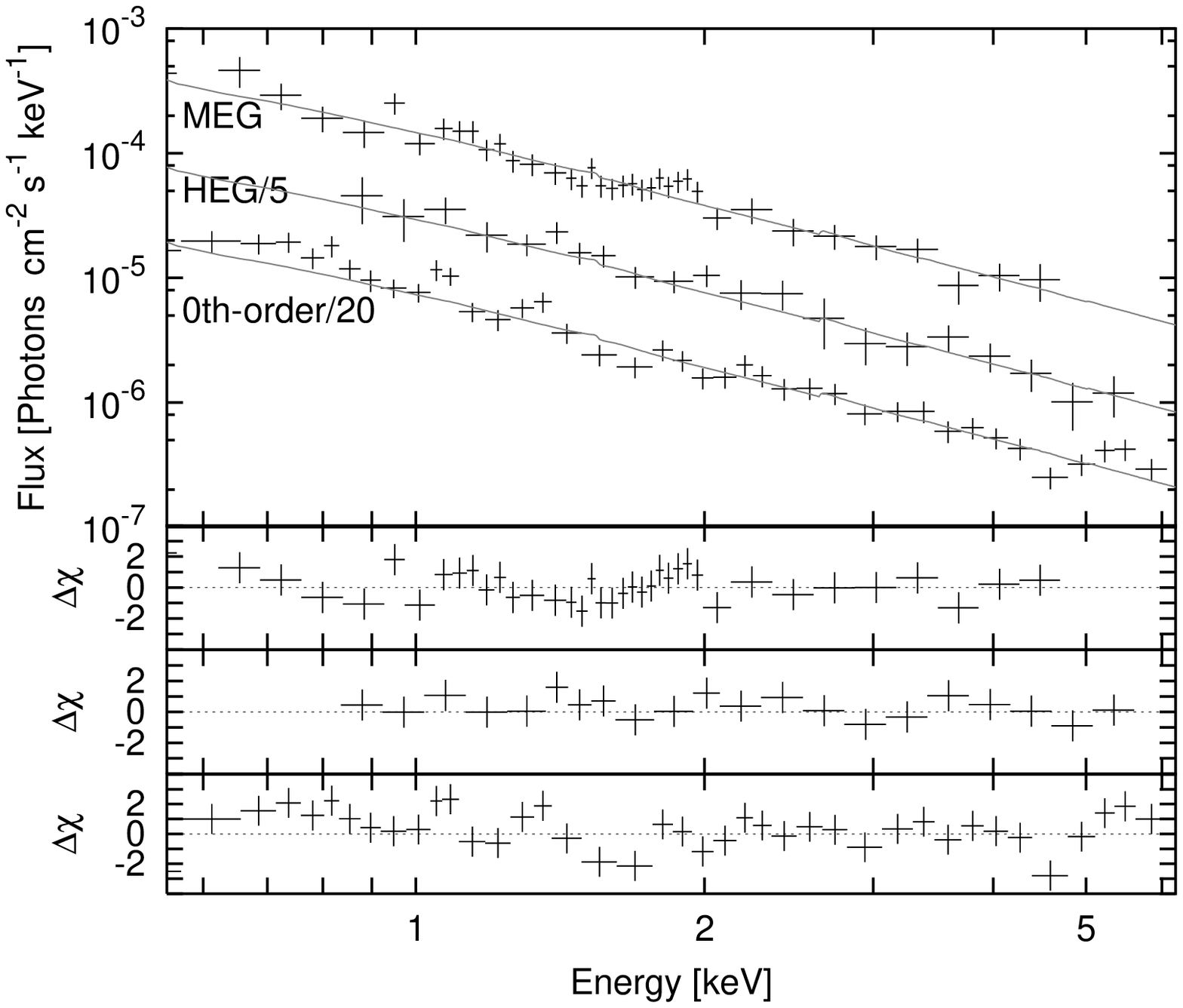}
\caption{
The $\pm1$ order HETGS (HEG/MEG) data for GRB~021004
are very well fit by an absorbed
power-law (Section \ref{sec:spectral_fitting}).  The 0th-order data
(S/N $\ge 5$ per bin) are also fit well by this model.
Here we divide the HEG data by a factor of 5 and the 0th-order data
by a factor of 20 for ease of viewing.
}
\label{fig:1004spec}
\end{center}
\end{figure*}

\begin{figure*}
\begin{center}
\epsscale{0.8}
\plotone{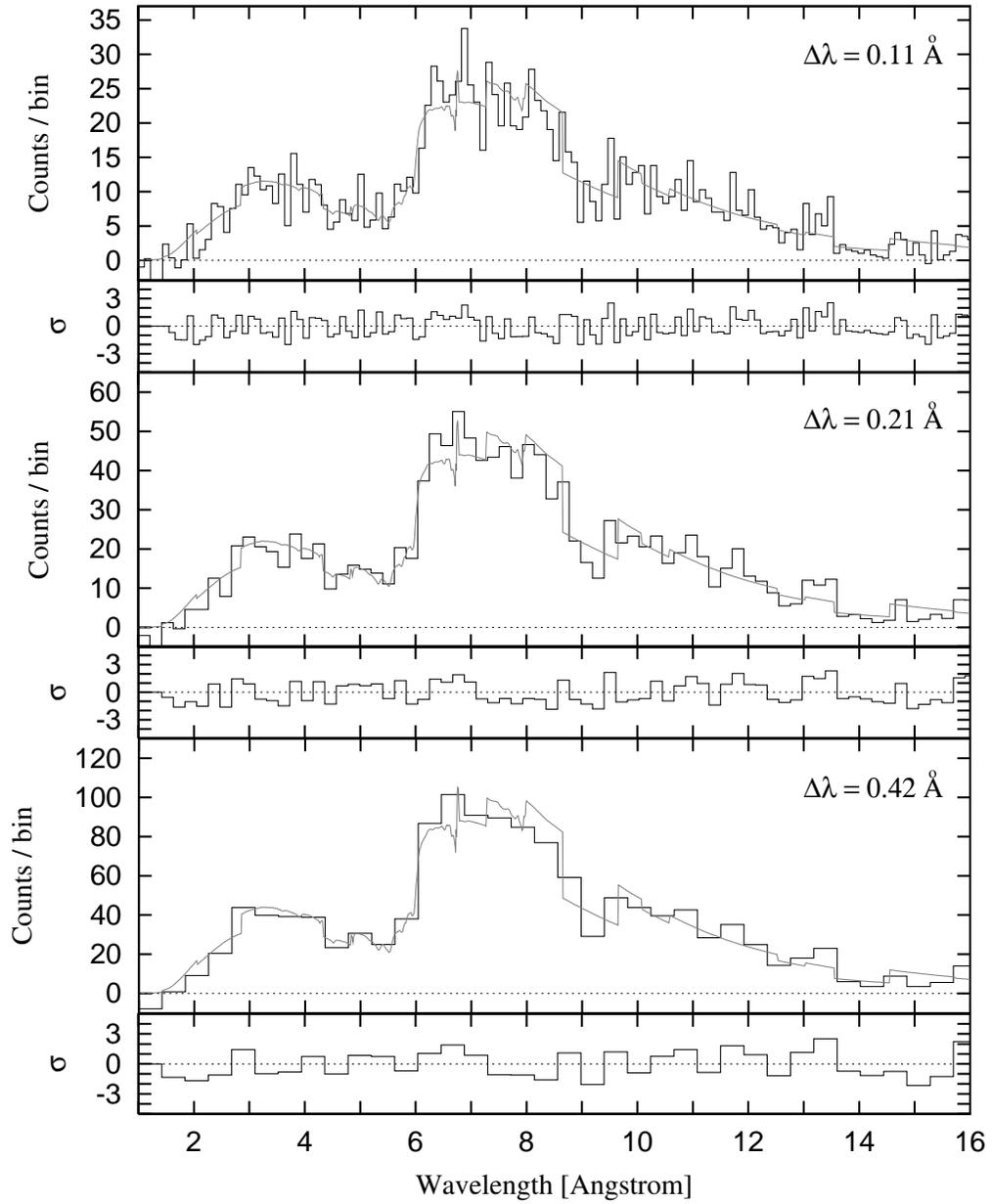}
\caption{
The best fit power-law model on top of the combined HETGS 
HEG+MEG data, at 3 binnings, for GRB~021004.  No highly
significant deviations from the model are observed.
}
\label{fig:allcts_1004}
\end{center}
\end{figure*}

\begin{figure*}
\begin{center}
\epsscale{1}
\plotone{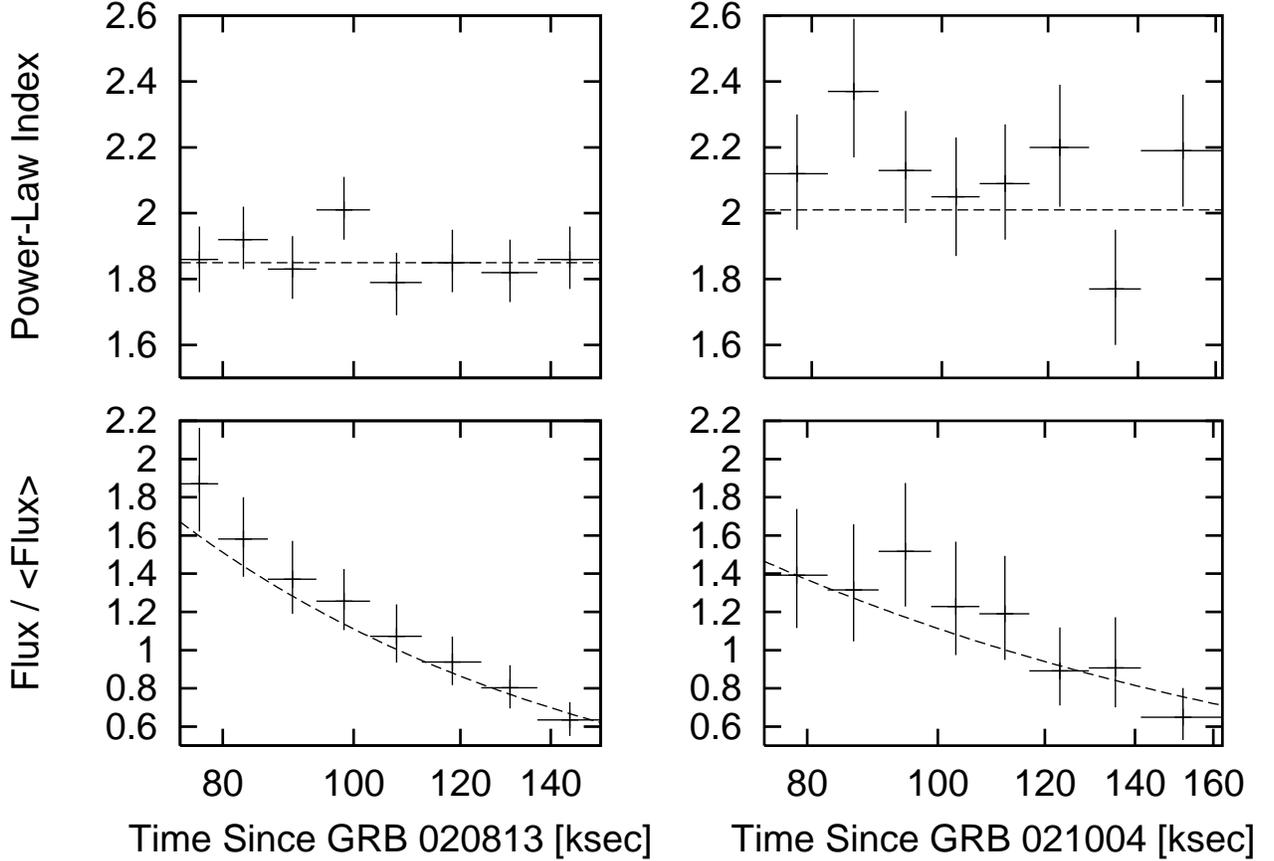}
\caption{
To look for spectral variability, the HETGS data for each GRB are 
divided into 8 regions of equal counts and
fit with absorbed power-laws by minimizing the Cash statistic \citep{cash79}.
The above plots show the power-law indeces $\Gamma$ and the (energy) fluxes
found for each fit, with the best-fit $\Gamma$'s derived for the entire
data set and the best-fit temporal fade found for the count rates overplotted 
as dashed lines.  The fluxes have been divided by
the mean fluxes for each observation reported in Section \ref{sec:observe}.
Error bars are 90\% confidence, determined from the 1-parameter
confidence intervals.
}
\label{fig:indxNnorm}
\end{center}
\end{figure*}

\begin{figure*}
\begin{center}
\plotone{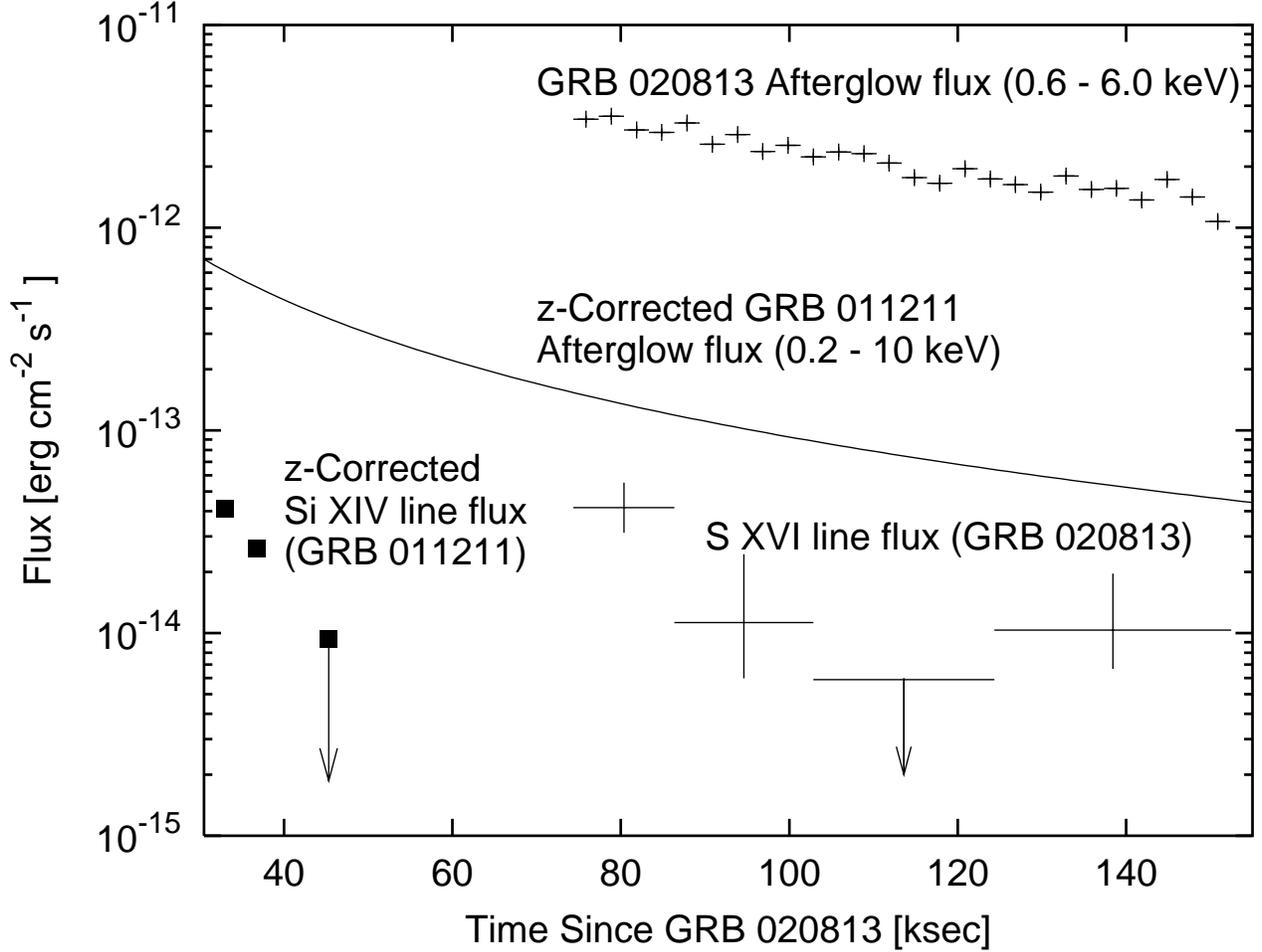}
\caption{
The points on the left show the Si XIV K$\alpha$ line 
flux for GRB~011211 from R02, corrected to the redshift of GRB~020813.
The integrated afterglow flux for GRB~012111 is also plotted and
extrapolated.  We employ a cosmology
with $\Omega_m=0.3$, $\Omega_{\Lambda}=0.7$, and $h=0.65$. 
At a later time ($\sim 50$ ksec later in the GRB frame), we measure 
a similar line flux.  However, the GRB~020813 afterglow flux is
approximately an order of magnitude higher that that of GRB~011211.  On the 
right, we plot the measured line flux from S XVI K$\alpha$, and the 
afterglow flux, versus time since GRB 020813.  
Error bars and the one upper limit are at 1$\sigma$ confidence.  
}
\label{fig:813line}
\end{center}
\end{figure*}

We see three reasons, which arise from complexities in our modeling,
to be cautious 
in interpreting our results as evidence for a persistent thermal component.
First, we observe lines with best fit widths an order of magnitude 
larger than the thermal line widths.  We account for this in the thermal 
model by introducing a turbulence velocity of $2000$ km/s.  It is also 
possible that this broadening is kinematical.  
Second, the best-fit CIE plasma model has marginally dissimilar abundances 
for the multiple-$\alpha$ elements (Figure \ref{fig:813abund}) and very low 
abundances for the Fe group elements relative to the multiple-$\alpha$ 
elements (Figure \ref{fig:813abund}, Figure \ref{fig:813fits} Inset).
This would require an unusual chemical composition for the emitting medium, a 
mechanism beyond the APED model for suppressing emission from metals other 
than Si and S, or the addition of a power-law continuum component to
the model as discussed in Section \ref{sec:spectral_fitting}. 
Third, the data do not tightly constrain the relative contributions to 
the continuum from the CIE plasma and from this possible power-law component.
Indeed, the continuum appears non-thermal, as would be expected for the 
X-ray afterglow of the burst.  

Though these considerations do not rule out a thermal model, reflection 
models may more naturally describe a spectrum with broad lines 
superposed on a power-law
continuum (see e.g. \citet{ballantyne01}, \citet{vietri01}).   In these 
models, the line broadening is due to Compton scattering.  If the electron 
temperature is similar to the line energy, we expect $\sigma_E/E \sim 
(kT/m_ec^2)^{1/2} \sim 0.05$ (at $1.3$ keV) for a single scattering.  
Because this is already a factor $\sim 5$ times larger than the observed 
widths, we require the reflector to not be highly ionized.  Consulting 
Figure 2 of Lazzati, Ramirez-Ruiz, \& Rees (2002), it appears that our
observed dissimilarity in line luminosities (Table \ref{table:abund},
Figure \ref{fig:813fluxes}) may further reinforce this conclusion.
Their reflection models with ionization parameter $\xi$ in the range 
$10<\xi<100$ can produce Si and S line luminosities which are an order 
of magnitude greater than luminosities of lines from Ar, Ca, and Fe.  
In the case of GRB~011211, Lazzati et al. (2002) argue that the large
line equivalent widths rule out nearby reprocessor models.  This
is not the case for GRB~020813, and the line duration may be set
by a delayed energy injection rather than by the geometry of the
reprocessing material (see \cite{reesnmeszaros00}, \citet{meszarosnrees01}).
In this case, we would not need a two-step (SN then GRB) explosion.

In conclusion, we detect
lines from an over-abundance of light metals characteristically 
produced in massive stars during pre-supernova nucleosynthesis 
in the X-ray afterglow of GRB~020813.  The long S~XVI line duration and 
narrow line widths can be explained via the geometry of the emitting material,
and this would likely require a two-step explosion as in the supranova
\citep{vietri99} scenario.  The time delay between the SN and GRB is
$\gtrsim 2$ months in this picture.  In contrast, our weak
detection of a Ni feature, with no detection of Fe (or Co), suggests
a very short time delay ($\lessim$ 1 week) between the SN and the GRB.
As discussed in \citet{woosely02}, it is not likely possible to observe
a GRB through a supernova remnant this young;
the Ni feature may indicate that the geometric picture is wrong.
A thermal model adequately fits the full 
data set.  A bright Fe (or Ni or Co) feature is not observed, 
however, and this is perhaps the strongest reason to favor a reflection 
model, where an Fe line can be quenched by Auger auto-ionization 
(Lazzati et al. 2002).

\acknowledgments

We thank Harvey Tananbaum for his generous allocation of Director's 
Discretion Time for these observations.  This research was supported in part 
by NASA contract NASW-4690.

\clearpage

\clearpage

\begin{table}[h]
\begin{center}
\begin{tabular}{cccc}
\hline
Element & Abund. & Flux & EW [eV] \\
\hline
 Mg &  4.4                       & 0.5 & 13 \\
 Si & 5.8 & 1.9 & 53 \\
  & ($2.9_{-1.6}^{+2.9}$) & ($1.1\pm 0.8$) & ($30 \pm 23$) \\
 S  &  4.3 & 2.3 & 72  \\
   &  ($2.9^{+1.4}_{-2.0}$) & ($1.6\pm 0.7$) & ($50 \pm 22$) \\
 Ar &  1.0                       & 0.2 & 9 \\
 Ca &  1.8                       & 0.4 & 15 \\
 Fe &  0.1                       & 1.5 & 100 \\
 Ni &  4.4        & 1.5 & 110 \\
  &  ($1.8^{+2.6}_{-1.8}$)        &  $1.0^{+0.5}_{-1.0}$ & $73^{+37}_{-73}$ \\
\hline
\end{tabular}
\caption{\noindent
GRB~020813 line emission upper limits at 90\% confidence for
APED model elemental 
abundances, H-like K$\alpha$ line fluxes ($\times 10^{-14}$ erg 
cm$^{-2}$ s$^{-1}$), 
and rest-frame H-like K$\alpha$ line equivalent widths.  Values
determined from power-law plus 
Gaussian fits. The observer-frame line widths were
set via $dE/E=v_t/c\approx 0.0067$.  Line centers are set
with $z=0.99$.
}
\label{table:abund}
\end{center}
\end{table}

\end{document}